%% file: main.tex
\documentclass[sigconf]{aamas}

\usepackage{amsmath}
\usepackage{mathtools}
\usepackage{booktabs}
\usepackage{graphicx}
\usepackage{subcaption}

\setcopyright{none}
\acmConference[arXiv]{arXiv preprint}{2026}{Online}{}
\copyrightyear{2026}
\acmYear{2026}
\acmDOI{}
\acmPrice{}
\acmISBN{}
\acmSubmissionID{}

\title[BARD-MARL]{BARD-MARL: Byzantine-Agent Detection for Learned Communication in Multi-Agent Reinforcement Learning}

\author{Almond Kiruthu Murimi}
\affiliation{
  \institution{Carnegie Mellon University}
  \department{Department of Electrical and Computer Engineering}
  \city{Pittsburgh}
  \state{Pennsylvania}
  \country{United States}}
\email{amurimi@andrew.cmu.edu}

\begin{abstract}
Learned communication improves coordination in cooperative multi-agent
reinforcement learning, but it also creates a trust problem: a trained policy
may route information through agents that have become faulty or adversarial.
This paper studies Byzantine-agent detection for learned-communication MARL in
adaptive traffic signal control. We propose \bard{}, a post-hoc diagnostic layer
on top of \bayesg{} \cite{duan2025bayesg}, which is used as an attributed
communication substrate rather than as a contribution of this paper. \bard{}
combines two agent-level evidence streams: policy-graph features extracted from
state-action trajectories and Bayesian trust statistics computed from
\bayesg{} latent mask probabilities. Across fixed-action, observation-flip,
random-noise, and coordinated attacks in SUMO traffic grids, the results show
that these signals are complementary rather than universally dominant. On a
25-agent grid, \bard{} reaches 0.843 AUC-ROC under a 10\%
observation-flip attack, while policy-graph-only detection reaches 0.917
AUC-ROC under a 10\% coordinated attack. On a 100-agent grid, the unified
\bard{} variant reaches 0.982 AUC-ROC for both 10\% fixed-action and 10\%
coordinated attacks. The study shows that learned communication policies expose
useful diagnostic evidence, but credible resilience claims require
attack-specific ablations and explicit separation between coordination,
detection, and mitigation.
\end{abstract}


\newcommand{\BibTeX}{\rm B\kern-.05em{\sc i\kern-.025em b}\kern-.08em\TeX}
\newcommand{\bard}{\textsc{BARD-MARL}}
\newcommand{\bayesg}{\textsc{BayesG}}

\newcommand{\iforest}{Isolation Forest}
\newcommand{\byz}{\mathcal{B}}
\newcommand{\agents}{\mathcal{N}}

\begin{document}

\raggedbottom

\pagestyle{fancy}
\fancyhead{}

\maketitle

\input{sections/01_introduction}
\input{sections/02_related_work}
\input{sections/03_problem_setup}
\input{sections/04_method}
\input{sections/05_experiments}
\input{sections/06_results}
\input{sections/07_discussion}
\input{sections/08_conclusion}

\appendix
\input{sections/appendix}

\bibliographystyle{ACM-Reference-Format}
\bibliography{references}

\end{document}

%% file: sections/01_introduction.tex
\section{Introduction}
\label{sec:introduction}

Communication is now a central design choice in cooperative multi-agent
reinforcement learning (MARL). Recent surveys of communication-based multi-agent
deep RL frame communication as a mechanism for improving coordination under
decentralized control. They also emphasize that messages can broaden each
agent's partial view of the environment and support collaboration
\cite{zhu2024commmarl}. Adaptive
traffic signal control is a natural example: an intersection can make better
phase decisions when it knows how queues and flows are evolving at neighboring
intersections, and large traffic grids require these local decisions to remain
coherent over time \cite{bokade2023representational}. In simulation platforms
such as SUMO \cite{lopez2018sumo}, communication-based MARL has therefore
become a practical testbed for studying coordination under partial observability.

The same communication channel creates a trust problem. A learned MARL policy
usually assumes that neighboring agents are cooperative, or at least that their
messages and actions are generated by the same training-time process. That
assumption is fragile in deployed infrastructure. Recent work on connected
adaptive traffic signal control shows that wireless traffic-control interfaces
increase the attack surface and can be manipulated to induce congestion
\cite{irfan2022atscattack}. In cooperative MARL more broadly, Byzantine failures
allow an agent to take arbitrary worst-case actions because of malfunction or
adversarial control \cite{li2024bardec}. The task studied in this paper is
therefore agent-level Byzantine detection for learned communication in MARL:
given a trained communication-based traffic-control policy and rollouts from a
network containing faulty or malicious agents, identify which agents should no
longer be trusted.

This task is difficult for three reasons. First, the performance signal in
traffic control is highly coupled: a low reward at one intersection may reflect
congestion propagated from elsewhere rather than local malicious behavior.
Second, modern communication policies do not always expose a fixed graph that can
be inspected directly. \bayesg{} \cite{duan2025bayesg}, for example, learns
sparse, context-dependent ego-graphs through Bayesian variational inference, so
the communication structure itself changes with local observations. Third,
Byzantine behavior does not have a single signature. A fixed-action attacker may
collapse its policy to one traffic phase, an observation-flip attacker may
corrupt the inputs that drive communication, and coordinated attackers may act
consistently enough to look less uncertain than isolated faults. A useful
detector must therefore distinguish behavioral abnormality from ordinary traffic
variation while remaining sensitive to different attack mechanisms.

Recent communication-based MARL work primarily answers a coordination question:
which information should an agent exchange, attend to, or route in order to
improve task performance? This framing appears in surveys of communication
design choices \cite{zhu2024commmarl}, traffic-signal communication policies
\cite{bokade2023representational}, graph-attention information dissemination
\cite{galliera2024collaborative}, GNN-based resilient coordination
\cite{goeckner2024magec}, and Bayesian ego-graph inference for networked MARL
\cite{duan2025bayesg}. These methods are valuable because they improve
coordination, scalability, or robustness under disturbances. They do not, by
themselves, answer a separate diagnostic question: which specific neighbor has
become unreliable? For this paper, reward ranking and observation z-scores are
therefore treated as intuitive post-hoc baselines, not as literature-backed
Byzantine detectors; their limitations are evaluated empirically in
Section~\ref{sec:results}.

This paper proposes \bard{}, a Byzantine-aware detection layer for learned
communication in MARL. \bard{} builds on \bayesg{} \cite{duan2025bayesg} as a
baseline communication policy, not as a contribution of this paper. The key
observation is that \bayesg{} \cite{duan2025bayesg} already produces a Bayesian
uncertainty signal while training and executing its variational communication
model. \bard{} repurposes this signal as part of a trust score, then combines it
with structural features extracted from per-agent policy graphs. Each policy
graph summarizes the agent's observed state-action behavior through a
shared k-means state abstraction, following the broader idea that structured
policy abstractions can make learned policies inspectable
\cite{bastani2018viper,topin2019policygraphs}. Sparse or collapsed graphs
indicate agents whose behavior has become simpler or less context sensitive than
honest controllers. These features are fused with VAE-derived statistics and
scored with \iforest{} \cite{liu2008isolation} to produce agent-level anomaly
rankings.

The resulting study supports a more nuanced claim than ``uncertainty detects all
attacks.'' Bayesian uncertainty and policy-graph structure are complementary,
and their usefulness depends on the attack. The headline results in
Section~\ref{sec:results} show this directly: on a 5-by-5 SUMO traffic grid,
\bard{} reaches an AUC-ROC of 0.843 under a 10\% observation-flip attack, while
policy-graph-only detection reaches 0.917 AUC-ROC under a 10\% coordinated
attack. On a 10-by-10 grid with 100 agents, the unified \bard{} variant reaches
0.982 AUC-ROC under both 10\% fixed-action and 10\% coordinated attacks. These
results indicate that learned communication systems can expose useful internal
diagnostic signals, but only when those signals are interpreted together with
behavioral evidence.

\noindent\textbf{\emph{Original contributions.}} This paper makes three
contributions. It formulates Byzantine detection as a separate diagnostic
problem for communication-based MARL in adaptive traffic signal control, rather
than treating it as ordinary reward degradation. It introduces \bard{}, which
combines policy-graph behavioral structure with Bayesian trust signals derived
from a \bayesg{} \cite{duan2025bayesg} communication model. It provides a
multi-attack, multi-seed evaluation across 25-agent and 100-agent SUMO traffic
grids, including ablations that show when VAE uncertainty helps, when policy
graphs dominate, and why attack-specific analysis is necessary for credible
resilience claims. Section~\ref{sec:related_work} positions the work against
communication MARL and robust MARL; Section~\ref{sec:method} defines the
detection pipeline; Section~\ref{sec:experiments} describes the SUMO evaluation;
Section~\ref{sec:results} reports the main evidence; and the appendix provides
implementation and reproducibility details.

%% file: sections/02_related_work.tex
\section{Related Works}
\label{sec:related_work}

Communication-based MARL studies how agents should exchange information when
each agent observes only part of the environment. Recent surveys organize this
literature around who communicates, what is communicated, when communication
occurs, and how learned messages affect decentralized coordination
\cite{zhu2024commmarl}. In adaptive traffic signal control, representational
communication has been used to let traffic agents select message content and
recipients rather than communicating with all neighboring agents at all times
\cite{bokade2023representational}. Related graph-based approaches use graph
neural networks or graph attention to route information over dynamic networks,
including information dissemination and resilient multi-robot coordination
\cite{galliera2024collaborative,goeckner2024magec}. These methods primarily
answer a coordination question: which relations should an agent use to improve
task performance?

\bayesg{} \cite{duan2025bayesg} is closest to the communication setting of this
paper. It learns sparse, context-aware ego-graphs for networked MARL through
Bayesian variational inference, allowing each agent to sample a latent
communication mask over its local neighborhood. In this paper, \bayesg{} is the
baseline and substrate, not a contribution. Differently from learned
communication methods that optimize which relations improve coordination,
\bard{} asks whether the learned communication process also exposes evidence
that a specific neighbor has become Byzantine.

Byzantine robustness has been studied in multi-agent control and MARL. Early
resilient MARL work considers networked $Q$-learning with Byzantine agents and
gives convergence conditions under graph robustness assumptions
\cite{xie2021resilience}. More recent work formulates Byzantine robust
cooperative MARL as a Bayesian game, where agents reason over posterior beliefs
about adversarial types and learn policies that remain useful under worst-case
perturbations \cite{li2024bardec}. Complementary theory studies decentralized
policy evaluation under Byzantine attacks and shows impossibility limits for
recovering the normal-agent value objective under heterogeneous local rewards
\cite{hairi2024hardness}. These works establish that Byzantine behavior is a
real obstacle for cooperative learning and distributed evaluation. They mainly
ask how to learn or evaluate policies despite adversaries; \bard{} instead asks
whether rollouts and internal communication signals can identify which agents
are unreliable after a communication-based policy has been trained. This
distinction matters for adaptive traffic signal control, where connected
traffic infrastructure can be attacked in ways that induce congestion while
appearing locally plausible \cite{irfan2022atscattack}.

Policy extraction and policy-level explanation provide tools for inspecting the
behavior of learned RL policies. VIPER trains decision-tree policies from a
neural policy and value information so that learned behavior can be verified
with structured representations \cite{bastani2018viper}. Abstracted Policy
Graphs summarize policies as Markov chains over abstract states, making
sequences of decisions explainable rather than explaining only isolated actions
\cite{topin2019policygraphs}. \bard{} uses this idea for detection rather than
explanation alone. The policy-graph features in this paper summarize each
traffic agent's observed state-action transitions through graph size, transition
structure, action entropy, and dominant-action frequency. The gap addressed here
is not whether a policy can be explained, but whether behavioral structure can
help distinguish honest agents from Byzantine agents whose actions become
sparse, repetitive, or inconsistent with the honest population.

Unsupervised anomaly detection supplies the final tool used in this paper.
\iforest{} isolates anomalous points through random recursive partitioning and
is widely used because it can score compact feature vectors without supervised
labels \cite{liu2008isolation}. Recent work extends this family to
set-structured network anomalies, showing that isolation-based detectors can be
adapted when observations are naturally grouped rather than independent points
\cite{djidjev2024siforest}. Byzantine detection has also been studied in
multi-robot systems, for example by comparing submitted visual observations and
using distributed ledger mechanisms to build trust under decentralization
\cite{salimpour2022vision}. \bard{} differs in both signal and domain. It does
not assume external image comparison, hand-labeled trust reports, or a separate
ledger. Instead, it detects suspicious traffic-control agents using two signals
already available from the learned MARL system: behavioral policy-graph
structure and Bayesian uncertainty from the \bayesg{} communication model
\cite{duan2025bayesg}. Taken together, prior work provides learned
communication mechanisms, robust MARL formulations, policy abstraction tools,
and anomaly detectors; \bard{} combines these threads for the narrower
diagnostic problem of identifying Byzantine agents in learned-communication
MARL for adaptive traffic signal control.

%% file: sections/03_problem_setup.tex
\section{Problem Setup}
\label{sec:problem}

We consider a cooperative adaptive traffic signal control problem with agents
$\agents=\{1,\ldots,n\}$, where each agent controls one intersection. At time
$t$, agent $i$ observes local traffic state $o_i^t \in \mathcal{O}_i$, selects a
traffic phase action $a_i^t \in \mathcal{A}_i$, and receives reward $r_i^t$.
The environment return is a global traffic-control objective, implemented in
SUMO as a congestion-sensitive reward. A communication-based policy lets agent
$i$ condition its action not only on $o_i^t$, but also on information routed
from neighboring agents. In our experiments this communication substrate is
\bayesg{} \cite{duan2025bayesg}, which learns a latent ego-graph over local
neighbors.

A hidden Byzantine subset $\byz \subset \agents$ is selected at evaluation time.
For honest agents, observations and actions are unchanged. For Byzantine agents,
an attack operator can corrupt the observation consumed by the policy and/or the
action executed in the environment:
\begin{equation}
  \tilde{o}_i^t =
  \begin{cases}
    A_o(o_i^t), & i \in \byz,\\
    o_i^t, & i \notin \byz,
  \end{cases}
  \qquad
  \tilde{a}_i^t =
  \begin{cases}
    A_a(a_i^t), & i \in \byz,\\
    a_i^t, & i \notin \byz.
  \end{cases}
\end{equation}
The implemented attacks instantiate $A_o$ and $A_a$ as fixed-action control,
observation flipping, random observation noise, and coordinated phase control.
The detector never observes $\byz$ directly.

After a rollout of length $T$, the detector receives per-agent trajectories
$\tau_i=\{(o_i^t,a_i^t,r_i^t,o_i^{t+1})\}_{t=1}^{T}$ and, when available, the
latent communication mask probabilities produced by \bayesg{} during the same
rollout. The detection task is to assign each agent a suspicion score
$s_i \in \mathbb{R}$, where larger values indicate stronger evidence of
Byzantine behavior, and to return a flagged set
$\widehat{\byz} \subseteq \agents$. In controlled experiments, the assumed
contamination rate determines the number of flagged agents; in deployment, the
same scores could be thresholded. Detection quality is evaluated against the
hidden set $\byz$ using precision, recall, F1, and AUC-ROC. Mitigation is
treated as a downstream response: given $\widehat{\byz}$, the policy can be
rerun with latent communication edges to flagged agents severed.

%% file: sections/04_method.tex
\section{BARD-MARL}
\label{sec:method}

\begin{figure*}[t]
  \centering
  \makebox[\textwidth][c]{%
    \includegraphics[width=1.08\textwidth]{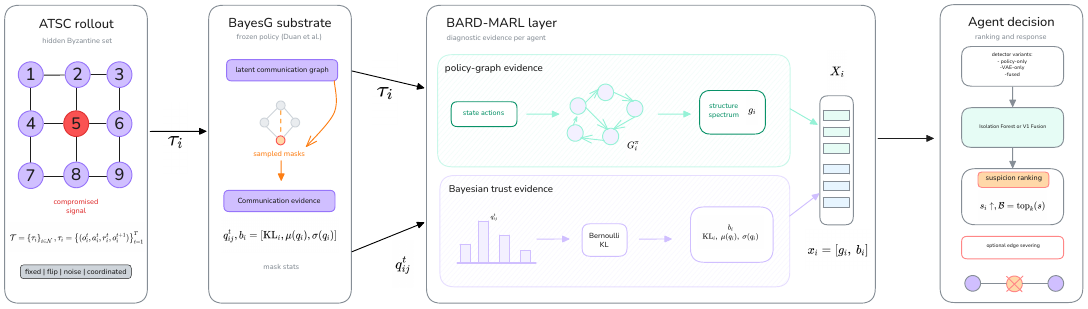}}
  \caption{\textbf{\bard{} overview.} A trained \bayesg{} communication policy
  \cite{duan2025bayesg} produces rollouts, latent masks, and Bayesian
  uncertainty signals in a traffic grid containing hidden Byzantine agents.
  \bard{} extracts two agent-level evidence streams: policy-graph behavioral
  structure from trajectories and VAE-derived trust statistics from the
  communication model. The detector scores each agent with policy-only,
  VAE-only, or fused features; optional mitigation then severs latent
  communication edges from flagged agents.}
  \label{fig:method-overview}
  \Description{A pipeline diagram showing traffic-control rollouts under attack, BayesG latent communication, BARD-MARL evidence extraction, agent-level anomaly scoring, and optional mitigation by edge severing.}
\end{figure*}

\bard{} is a post-hoc diagnostic layer for a trained MARL communication policy.
It does not retrain \bayesg{} and does not claim the learned
communication mechanism as a contribution. Instead, it uses two signals already
available during evaluation: the observed state-action behavior of each traffic
agent and the latent mask probabilities produced by \bayesg{}
\cite{duan2025bayesg}. Figure~\ref{fig:method-overview} summarizes the
pipeline.

\subsection{Policy-Graph Evidence}

For each agent, \bard{} extracts a compact policy graph from trajectory data,
following the broader idea that learned RL policies can be inspected through
structured policy representations \cite{bastani2018viper,topin2019policygraphs}.
First, observations from all agents are standardized and clustered into a shared
state vocabulary with $K=8$ k-means clusters \cite{macqueen1967kmeans}. Let
$c_i^t$ denote the cluster assigned to agent $i$'s observation at time $t$.
\bard{} defines a policy-graph node as the abstract state-action pair
$v_i^t=(c_i^t,a_i^t)$ and adds a directed transition
$v_i^t \rightarrow v_i^{t+1}$ for each consecutive trajectory step. Edge weights
are normalized transition frequencies, so the graph records how the agent moves
between abstract decision modes rather than only how often each action appears.

Given the directed policy graph $G_i^{\pi}=(V_i,E_i,W_i)$, \bard{} computes
structural features including $|V_i|$, $|E_i|$, density, average in-degree,
standard deviation of in-degree, average out-degree, number of strongly
connected components, and transition entropy
\begin{equation}
  H_i^{\mathrm{trans}} =
  - \sum_{(u,v)\in E_i} w_{uv}\log(w_{uv}+\epsilon).
\end{equation}
It also computes spectral features from the symmetrized weighted adjacency
$A_i^{\mathrm{sym}}=(A_i + A_i^\top)/2$. With
$L_i=D_i-A_i^{\mathrm{sym}}$, the feature vector uses the non-trivial Laplacian
eigenvalues $\lambda_2,\ldots,\lambda_6$, skipping the always-zero trivial
eigenvalue. Finally, \bard{} appends trajectory-level behavioral statistics:
mean reward, reward standard deviation, action entropy, dominant-action
frequency, observation mean, and observation standard deviation. The resulting
policy-graph feature vector is denoted $g_i$.

\subsection{Bayesian Trust Evidence}

\bayesg{} samples latent communication masks as part of its Bayesian ego-graph
inference mechanism \cite{duan2025bayesg}. During evaluation, the trained policy
produces mask probabilities $q_{ij}^t$ for agent $i$'s communication edge to
neighbor $j$. \bard{} converts these probabilities into a per-agent Bernoulli KL
signal against the sparse mask prior $\lambda=0.2$:
\begin{equation}
  k_i =
  \frac{1}{T}\sum_{t=1}^{T}\sum_{j\in \mathcal{N}_i}
  q_{ij}^t \log\frac{q_{ij}^t}{\lambda}
  + (1-q_{ij}^t)\log\frac{1-q_{ij}^t}{1-\lambda}.
\end{equation}
The KL term measures how far agent $i$'s learned communication behavior moves
from the sparsity prior over the rollout. \bard{} also records the mean and
standard deviation of the same mask probabilities. The Bayesian evidence vector
is therefore $b_i=[k_i,\mu(q_i),\sigma(q_i)]$.

\subsection{Detector Variants}

\bard{} evaluates three related detector variants, all producing a suspicion
ranking over agents. The policy-graph-only variant standardizes the vectors
$g_i$ and applies \iforest{} \cite{liu2008isolation}, using the negative
Isolation Forest score as suspicion. This ablation tests whether behavioral
structure alone is sufficient.

The first full variant, \bard{}-V1, linearly fuses policy-graph anomaly evidence
with the Bayesian KL signal. Let $\tilde{k}_i$ be the min-max normalized KL
value and let $\tilde{u}_i$ be the min-max normalized anomaly score
$-\mathrm{IF}(g_i)$. The trust score is
\begin{equation}
  T_i = \alpha(1-\tilde{k}_i) + \beta(1-\tilde{u}_i),
  \qquad \alpha=\beta=0.5,
\end{equation}
and suspicion is $s_i=1-T_i$. Lower trust therefore corresponds to stronger
Byzantine evidence. Setting $\alpha=1,\beta=0$ gives the VAE-only ablation.

The second full variant, \bard{}-V2, avoids manual linear fusion. It concatenates
the two evidence streams,
\begin{equation}
  x_i = [g_i, b_i],
\end{equation}
standardizes the resulting feature matrix across agents, and fits a single
\iforest{} \cite{liu2008isolation}. This variant lets the anomaly detector learn
whether policy-graph structure, Bayesian uncertainty, or their interaction is
most discriminative for the current attack.

\subsection{Response Mechanism}

The core contribution of \bard{} is detection. As a downstream response, the
flagged set $\widehat{\byz}$ can be passed back into the \bayesg{} policy and
used to alter the attention mask before graph-message aggregation. For each
local attention matrix, \bard{} zeros the row and column entries corresponding
to flagged neighbors while preserving diagonal self-loops. This severs latent
communication edges to suspicious agents without retraining the policy or
changing the learned model parameters. Detection and mitigation are reported
separately so that an accurate detector is not conflated with the downstream
traffic-control effect of severing communication edges.

%% file: sections/05_experiments.tex
\section{Experimental Setup}
\label{sec:experiments}

All experiments use adaptive traffic signal control in SUMO
\cite{lopez2018sumo}. Each traffic light is an agent and each agent controls
one intersection. The primary setting is a 5-by-5 grid with 25 agents; a
10-by-10 grid with 100 agents evaluates whether the detector remains useful
when the number of agents increases. The communication substrate is a trained
\bayesg{} policy \cite{duan2025bayesg}. We use \bayesg{} as an attributed
baseline and source of latent mask probabilities; the experiments do not retrain
\bayesg{} as part of \bard{}.

\subsection{Attacks and Evaluation Protocol}

Byzantine agents are introduced only at evaluation time. We evaluate four
implemented attacks: fixed-action control, where a compromised signal always
executes the same phase; observation flipping, where reported queue information
is inverted; random observation noise, where Gaussian sensor noise is injected;
and coordinated phase control, where multiple compromised intersections target
the same phase. Byzantine fractions are 10\%, 20\%, and 30\%. A 0\% condition is
used for clean coordination reference, not for AUC-ROC because only one class is
present.

For each attack condition, the detector receives the same evidence available to
the trained policy during rollout: per-agent trajectories and, for VAE-based
variants, \bayesg{} mask probabilities. The controlled experiments use the
known Byzantine fraction as the contamination rate, so detectors return the
corresponding top-ranked agents. This evaluates ranking quality under a fixed
operating point; threshold selection for deployment is left for future work.

\subsection{Baselines and Ablations}

The detection baselines are random flagging, reward-rank detection,
observation z-score detection, Bayesian-belief detection, and message
consistency detection. Random flagging is the chance reference. Reward rank
flags the agents with lowest cumulative reward. Observation z-score uses
statistical deviation in observations. The Bayesian-belief and message
consistency baselines are included where their required signals are available.

The ablations isolate the two evidence streams in \bard{}. The policy-graph
variant uses graph, spectral, and behavioral features without VAE mask
statistics. The VAE-only variant uses the Bernoulli KL-derived trust signal
without policy-graph features. \bard{}-V1 linearly fuses the VAE trust signal
with policy-graph anomaly scores, while \bard{}-V2 fits one \iforest{} on the
concatenated policy-graph and VAE feature vector.

\subsection{Metrics and Provenance}

Detection performance is reported with precision, recall, F1, and AUC-ROC.
AUC-ROC is the primary ranking metric because all detectors produce continuous
suspicion scores, while F1 reflects the fixed top-$k$ operating point implied by
the known Byzantine fraction. Reported detector values are aggregated over five
detection seeds for k-means and \iforest{} randomness when the frozen artifact
contains multi-seed output. The canonical result source for this manuscript is
\texttt{final\_paper\_results/}; figures and tables in this paper are generated
from those JSON artifacts.

%% file: sections/06_results.tex
\section{Results}
\label{sec:results}

\input{tables/table_detection_auc}

\begin{figure*}[t]
  \centering
  \includegraphics[width=\textwidth]{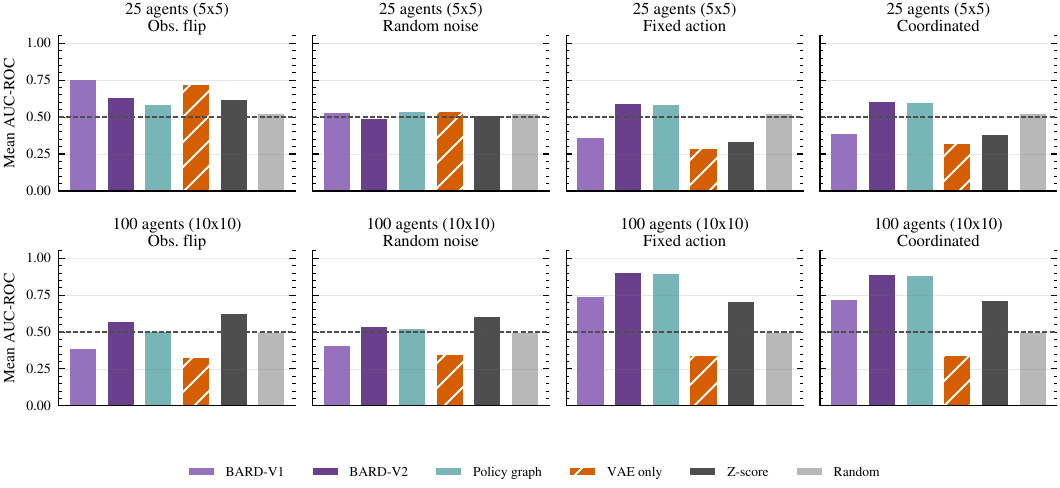}
  \caption{Mean AUC-ROC by attack and grid size. Each bar averages the 10\%,
  20\%, and 30\% Byzantine fractions for the corresponding attack. The dashed
  line marks random-chance ranking.}
  \label{fig:detection-auc-summary}
  \Description{A two-row bar chart comparing BARD variants, ablations, z-score, and random baselines across attacks on 25-agent and 100-agent traffic grids.}
\end{figure*}

Table~\ref{tab:detection-auc} and Figure~\ref{fig:detection-auc-summary} show
that Byzantine detection in learned-communication MARL is attack-dependent. On
the 25-agent grid, \bard{}-V1 is strongest on observation flipping, with mean
AUC-ROC 0.752 across Byzantine fractions. The policy-graph and unified
\bard{}-V2 variants are more stable on fixed-action and coordinated attacks,
where behavior structure carries more signal than VAE uncertainty alone. On the
100-agent grid, \bard{}-V2 and policy-graph-only detection become much stronger
for fixed-action and coordinated attacks, reaching mean AUC-ROC values above
0.88 for both attack families. Observation-flip and random-noise attacks remain
harder: simple z-score observation statistics are competitive there, which
suggests that those attacks leave stronger marginal observation signatures than
communication-structure signatures.

\input{tables/table_key_results}

\subsection{Detection Accuracy}

The high-signal conditions in Table~\ref{tab:key-results} illustrate the main
claim. Under 10\% observation flipping on the 25-agent grid, \bard{}-V1 reaches
0.843 AUC-ROC and 0.500 F1. Under 10\% coordinated attack on the same grid,
policy-graph-only detection reaches 0.917 AUC-ROC, showing that behavioral
structure can be more discriminative than VAE trust when coordinated agents act
confidently. On the 100-agent grid, \bard{}-V2 reaches 0.982 AUC-ROC for 10\%
fixed-action and 0.982 AUC-ROC for 10\% coordinated attacks, with F1 scores of
0.760 and 0.780 respectively.

These results do not support a universal detector-dominance claim. Instead, they
support a narrower and more useful claim: learned communication policies expose
diagnostic evidence, but the useful evidence stream depends on how the
Byzantine behavior manifests. VAE-derived trust is helpful for some
observation-mediated conditions, while policy-graph structure dominates when an
agent's decision dynamics collapse or become coordinated.

\subsection{Ablation and Scalability}

\begin{figure}[H]
  \centering
  \includegraphics[width=\columnwidth]{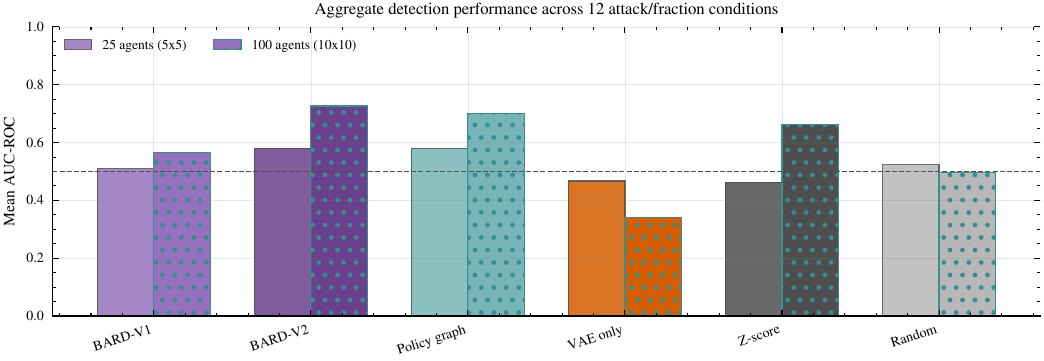}
  \caption{Aggregate AUC-ROC across the 12 attack/fraction conditions. The
  100-agent grid strengthens policy-graph and \bard{}-V2 evidence on structural
  attacks, while VAE-only detection weakens when used without behavioral
  features.}
  \label{fig:scalability-auc}
  \Description{A grouped bar chart comparing mean AUC-ROC on 25-agent and 100-agent grids for BARD variants, ablations, z-score, and random baselines.}
\end{figure}

Figure~\ref{fig:scalability-auc} aggregates across all four attacks and three
Byzantine fractions. \bard{}-V2 improves from 0.581 mean AUC-ROC on the
25-agent grid to 0.727 on the 100-agent grid. Policy-graph-only detection shows
a similar increase, from 0.579 to 0.701. In contrast, VAE-only detection drops
from 0.467 to 0.340, which indicates that the mask-probability KL signal is not
sufficient as a standalone detector at larger scale. This is consistent with
the method design: VAE statistics are a useful evidence stream, but \bard{} is
most reliable when the communication evidence is interpreted together with
policy-graph behavior.

\subsection{Mitigation as Auxiliary Evidence}

The core result of this paper is detection. The edge-severing response is
evaluated as an auxiliary mechanism because mitigation quality depends on both
detector accuracy and the downstream traffic-control effect of removing
communication edges. The frozen mitigation verdict marks the current mitigation
sweep as appendix-only: it does not pass the predefined success bars across all
conditions, and one random-noise condition shows statistically significant harm.
For that reason, the main claims do not rely on mitigation. The appendix reports
the mitigation sweep as preliminary evidence for future response design.

\newpage

%% file: tables/table_detection_auc.tex
\begin{table*}[t]
  \caption{Mean AUC-ROC across Byzantine fractions for each attack. Each cell averages the 10\%, 20\%, and 30\% Byzantine conditions; detector results use five detection seeds when available. Bold marks the best method within a grid and attack among methods with available results.}
  \label{tab:detection-auc}
  \centering
  \scriptsize
  \setlength{\tabcolsep}{4pt}
  \begin{tabular}{llccccc}
    \toprule
    Grid & Method & Obs. flip & Random noise & Fixed action & Coordinated & Mean \\
    \midrule
    25 agents (5x5) & BARD-V1 & \textbf{0.752} & 0.530 & 0.364 & 0.392 & 0.510 \\
     & BARD-V2 & 0.636 & 0.488 & \textbf{0.592} & 0.607 & 0.581 \\
     & Policy graph & 0.589 & \textbf{0.541} & 0.584 & 0.602 & 0.579 \\
     & VAE only & 0.719 & 0.538 & 0.289 & 0.324 & 0.467 \\
     & Z-score & 0.620 & 0.511 & 0.337 & 0.380 & 0.462 \\
     & Msg. consistency & 0.736 & N/A & 0.551 & \textbf{0.608} & 0.632 \\
     & Bayesian belief & 0.673 & N/A & 0.464 & 0.447 & 0.528 \\
     & Reward rank & 0.500 & 0.500 & 0.500 & 0.500 & 0.500 \\
     & Random & 0.523 & 0.523 & 0.523 & 0.523 & 0.523 \\
    \midrule
    100 agents (10x10) & BARD-V1 & 0.388 & 0.411 & 0.742 & 0.721 & 0.566 \\
     & BARD-V2 & 0.571 & 0.541 & \textbf{0.905} & \textbf{0.889} & 0.727 \\
     & Policy graph & 0.503 & 0.524 & 0.895 & 0.880 & 0.701 \\
     & VAE only & 0.331 & 0.348 & 0.340 & 0.341 & 0.340 \\
     & Z-score & \textbf{0.628} & \textbf{0.607} & 0.705 & 0.713 & 0.663 \\
     & Msg. consistency & N/A & N/A & N/A & N/A & N/A \\
     & Bayesian belief & N/A & N/A & N/A & N/A & N/A \\
     & Reward rank & 0.500 & 0.500 & 0.500 & 0.500 & 0.500 \\
     & Random & 0.497 & 0.497 & 0.497 & 0.497 & 0.497 \\
    \bottomrule
  \end{tabular}
\end{table*}

%% file: tables/table_key_results.tex
\begin{table}[H]
  \caption{Representative high-signal conditions used in the main text.}
  \label{tab:key-results}
  \centering
  \small
  \begin{tabular}{p{0.50\columnwidth}ccc}
    \toprule
    Condition & Method & AUC & F1 \\
    \midrule
    5x5 observation flip, 10\% & BARD-V1 & 0.843 & 0.500 \\
    5x5 coordinated, 10\% & Policy graph & 0.917 & 0.480 \\
    10x10 fixed action, 10\% & BARD-V2 & 0.982 & 0.760 \\
    10x10 coordinated, 10\% & BARD-V2 & 0.982 & 0.780 \\
    \bottomrule
  \end{tabular}
\end{table}

%% file: sections/07_discussion.tex
\section{Discussion}
\label{sec:discussion}

\bard{} is best understood as a diagnostic layer for learned communication, not
as a replacement for the underlying MARL controller. \bayesg{} provides the
trained communication policy and latent mask probabilities
\cite{duan2025bayesg}; \bard{} asks whether those signals, combined with
trajectory-derived policy graphs, can identify unreliable agents after training.
This separation is important because good coordination and good diagnosis are
not the same objective.

The attack-specific behavior is central to the result. Observation flipping and
random noise perturb the information entering the policy, so simple observation
statistics can remain competitive. Fixed-action and coordinated attacks change
the structure of behavior more directly, and the policy-graph evidence becomes
much stronger, especially on the 100-agent grid. The poor standalone
performance of VAE-only detection at 100 agents also clarifies the role of
Bayesian uncertainty: it is a useful feature, not a complete detector.

The main limitation is scope. The experiments use SUMO traffic grids and a
\bayesg{} communication substrate; they do not establish formal Byzantine
robustness guarantees, nor do they prove that the same feature stack transfers
unchanged to other MARL domains. The evaluation also uses the true Byzantine
fraction as the contamination rate. This is appropriate for controlled
comparison, but deployment would require threshold calibration or an explicit
prior over attack prevalence. Finally, edge severing is only a preliminary
response mechanism. It can improve some attacked conditions, but the current
sweep is not strong enough to claim reliable mitigation.

Despite these limitations, the results suggest a useful research direction.
Rather than treating learned communication as an opaque coordination module, we
can mine its internal signals and induced behavior for diagnostic evidence.
The strongest version of this idea is not a single trust score, but a
multi-signal detector whose ablations reveal which evidence stream is reliable
under each attack.

%% file: sections/08_conclusion.tex
\section{Conclusion}
\label{sec:conclusion}

This paper introduced \bard{}, a Byzantine-agent detection layer for
learned-communication MARL in adaptive traffic signal control. Built on top of
an attributed \bayesg{} substrate \cite{duan2025bayesg}, \bard{} combines
policy-graph behavioral structure with Bayesian trust statistics from latent
communication masks. The experiments show that these signals are complementary:
VAE-derived trust helps in some observation-mediated attacks, while
policy-graph structure is especially useful for fixed-action and coordinated
attacks at larger scale. The main takeaway is therefore not that one detector
wins universally, but that learned communication systems expose multiple
diagnostic signals that can be turned into agent-level Byzantine rankings.
Future work should study threshold calibration without known contamination
rates, stronger mitigation policies after detection, and transfer beyond traffic
signal control.

%% file: sections/appendix.tex
\section{Implementation Details}
\label{app:implementation}

\bard{} is implemented as a post-hoc evaluation layer. Policy graphs are built
from saved trajectory arrays containing observations, actions, rewards, and
next observations. The shared state vocabulary uses $K=8$ k-means clusters. For
each agent, graph features, Laplacian spectral features, and behavioral
statistics are concatenated before standardization and \iforest{} scoring.
VAE-derived features are computed from \bayesg{} mask probabilities saved during
the same rollout.

The frozen result artifacts use five detector seeds: 42, 123, 456, 789, and
1337. Evaluation rollouts use seeds 2000, 2010, 2020, 2030, and 2040 where
available. The canonical paper evidence is stored under the frozen result
directory. The paper figures and generated tables can be recreated with:
\begin{verbatim}
python scripts/generate_final_paper_assets.py \
  --results-root final_paper_results \
  --figures-dir paper/figures \
  --tables-dir paper/tables
\end{verbatim}

\section{Additional Results}
\label{app:additional-results}

\begin{figure}[H]
  \centering
  \includegraphics[width=\columnwidth]{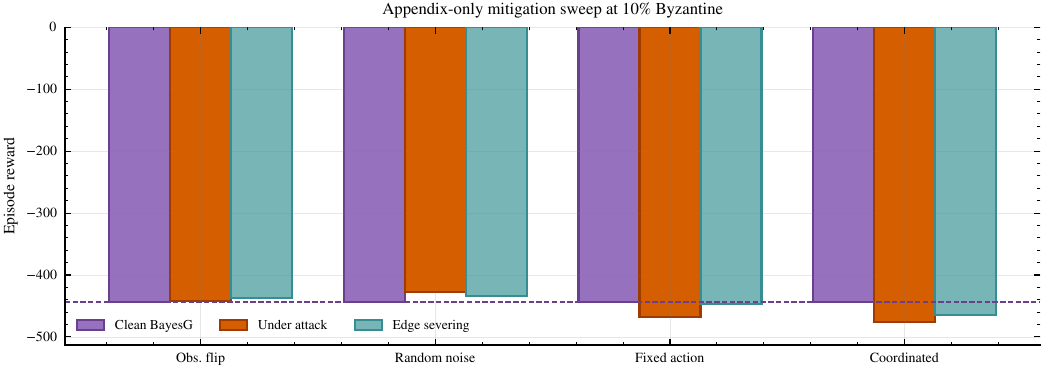}
  \caption{Appendix-only mitigation sweep at 10\% Byzantine. Edge severing is
  reported as auxiliary response evidence and is not used for the main
  detection claims.}
  \label{fig:mitigation-appendix}
  \Description{A grouped bar chart comparing clean BayesG, attacked, and edge-severed rewards under four attack types at 10 percent Byzantine agents.}
\end{figure}

Figure~\ref{fig:mitigation-appendix} summarizes the 10\% Byzantine mitigation
conditions. The full frozen verdict marks the mitigation branch as
appendix-only because the predefined success bars are not satisfied across all
conditions. This result is useful mainly as a warning: accurate detection is
necessary for response, but edge severing itself must be evaluated as a separate
control intervention.

\section{Reproducibility Checklist}
\label{app:reproducibility}

\input{tables/table_repro_artifacts}

The active attack names in the codebase are:
\begin{itemize}
  \item \texttt{fixed\_action}
  \item \texttt{observation\_flip}
  \item \texttt{random\_noise}
  \item \texttt{coordinated}
\end{itemize}
Earlier project notes used \texttt{random\_action} in a few places; the
manuscript and final artifacts use \texttt{random\_noise} to match the
implementation.

%% file: tables/table_repro_artifacts.tex
\begin{table}[H]
  \caption{Reproducibility artifacts checked for this preprint.}
  \label{tab:repro-artifacts}
  \centering
  \small
  \begin{tabular}{p{0.34\columnwidth}p{0.58\columnwidth}}
    \toprule
    Artifact & Status \\
    \midrule
    Frozen JSON results & 216 files under \texttt{final\_paper\_results/} \\
    Detection seeds & 5 detector seeds per reported condition where available \\
    Attack names & \texttt{fixed\_action}, \texttt{observation\_flip}, \texttt{random\_noise}, \texttt{coordinated} \\
    Figure assets & 13 files under \texttt{paper/figures/} before final compile \\
    Manuscript format & AAMAS 2026 style used for arXiv preprint \\
    \bottomrule
  \end{tabular}
\end{table}